%% file: main.tex
\documentclass[a4paper, amsfonts, amssymb, amsmath, reprint, showkeys, nofootinbib, superscriptaddress]{revtex4-2}
\usepackage[english]{babel}
\usepackage[utf8]{inputenc}
\usepackage[colorinlistoftodos, color=green!40, prependcaption]{todonotes}
\input{preamble}
\usepackage[pdftex, pdftitle={Article}, pdfauthor={Author}]{hyperref} 
\begin{document}
\title{A novel mechanism for spontaneous $\mathbf{E\times B}$ shear layer formation in scrape-off-layer plasmas}

\author{B Shanahan}
\email[Correspondence email address: ]{shanahan@psfc.mit.edu}

\affiliation{MIT Plasma Science and Fusion Center, 167 Albany St, Cambridge MA USA, 02139}

\author{E Rodriguez}   

    \affiliation{Max-Planck-Institut f\"ur Plasmaphysik, Telinstitut Greifswald, 17491 Greifswald, Germany}

\date{\today} 

\begin{abstract}
We present a novel mechanism for spontaneous shear-flow generation in the scrape-off layer (SOL) of magnetically confined plasmas, arising at abrupt changes in the sheath connection length $L_\parallel$. Such transitions occur naturally in the Wendelstein 7-X stellarator. Applying a critical balance treatment for interchange turbulence, we find that $L_\parallel$ sets the equilibrium pressure gradient and hence the radial electric field away from the transition, producing an $E\times B$ shear around the transition comparable to the turbulence growth rate. High-fidelity simulations are shown to be consistent with the theoretical prediction. 
\end{abstract}


\maketitle
A major challenge for fusion energy is the successful mitigation of heat and particle loads on plasma-facing components while maintaining a high performance plasma core. This can be achieved by inducing a transport barrier at the separatrix -- the interface between the closed-field lines in the core and the open-field lines in the scrape-off layer (SOL). This transport barrier can arise via a sheared radial electric field $E_x$, which drives an $E\times B$ flow that reduces transport~\cite{Biglari1990, Terry2000} by decorrelating turbulent eddies~\cite{Diamond1994, Hahm1995}. Identifying what sets $E_x$ in the SOL is therefore critical to the understanding of edge transport.

In the SOL of the Wendelstein 7-X (W7-X) stellarator \cite{Klinger2019}, strongly sheared poloidal flows have been recently observed~\cite{maragkoudakis2023interaction, Baek2024, Ballinger2026, Grulke2026}, where they affect edge transport~\cite{Killer2026} and advect turbulent filaments~\cite{Killershanahan2020, Zoletnik2019}. 

Crucially, the shear layers are located near abrupt transitions in the parallel connection length~\cite{Baek2024, Ballinger2026}, consistent with measurements in Heliotron J~\cite{Miyashita2024}. The mechanism responsible for these flows in W7-X has not, however, been identified.

In tokamaks, changes in connection length at the SOL entrance also give rise to strongly sheared $E\times B$ flows~\cite{Walkden2019,Walkden2022}, attributed to a strong temperature gradient driving the plasma sheath drop \cite[Eq.~(17.2)]{Stangeby}. In this Letter, we show that a radial electric field shear can be induced by a change in connection length without relying on such a mechanism. 

\begin{figure}
    \centering
    \includegraphics[width=\linewidth]{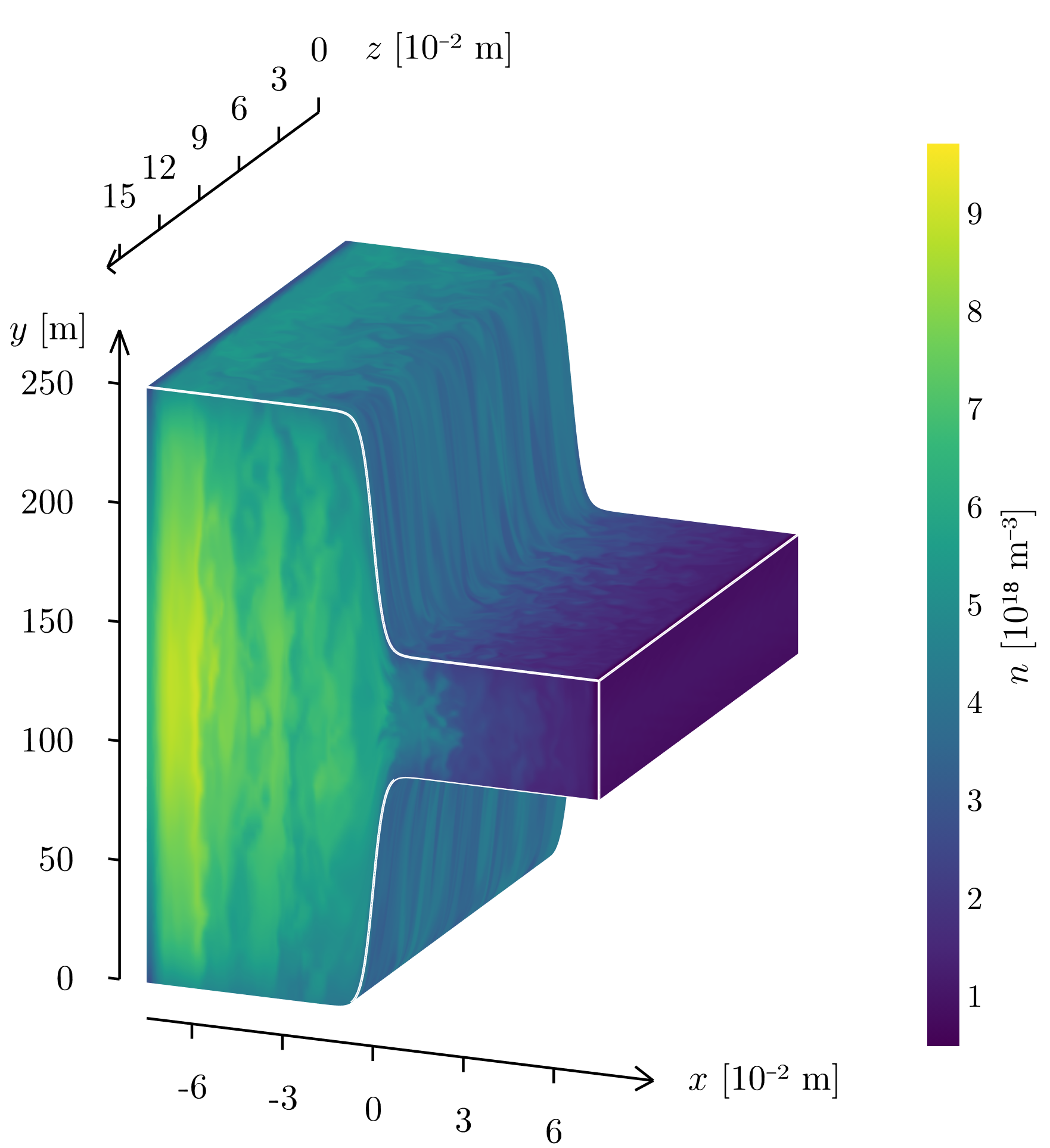}
    \caption{\textbf{Simulation domain.} An illustration of the simulated domain, showing a snapshot of the simulated density. The conducting walls are placed at the boundary of the $y$-domain.}
    \label{fig:gridsnap}
\end{figure}

To investigate the consequences of sharp transitions in the connection length, we perform drift-reduced two fluid turbulence simulations using an electrostatic Hermes-3 model~\cite{Dudson2023,Dudson2026} within the BOUT++ framework~\cite{Dudson2009,Dudson2015}. The simulations evolve the full density, pressure, momenta, and electrostatic potential in a simplified, curved, three-dimensional slab geometry, employing the Oberbeck--Boussinesq approximation for vorticity~\cite{Oberbeck1879, Gray1976, Bodi2011}. To model the connection-length transition (see Figure~\ref{fig:gridsnap}), we impose conducting-sheath
boundary conditions~\cite{Tskhakaya2005} at the field-aligned boundaries
\begin{equation}
    y = y_{\pm}(x) = y_c \pm \frac{1}{2}
    \left[L_0 - \Delta L\,\tanh\!\left(\frac{x}{\Delta x}\right)\right],
\end{equation}
where $y_c$ is the fixed midpoint of the domain, and $L_{\parallel,i}=L_0+\Delta L$ and $L_{\parallel,f}=L_0-\Delta L$
are the asymptotic connection lengths on the left and right, respectively. We perform a series of simulations spanning different magnitudes of the connection-length jump, fixing $L_{\parallel,i}=250~$m and changing $L_{\parallel,f}=50, 100, 150, 200, 250~$m.

The radial ($x$) and binormal ($z$) extents of the drift plane are held fixed ($L_x=L_z=15\,\mathrm{cm}$), but are subject to different boundary conditions.
The binormal direction is periodic, while the turbulence is driven by a localized source near the left ($x=-L_x/2$) boundary of the radial domain, mimicking particle and energy input from the core plasma. The right ($x=L_x/2$) boundary is set to have zero-gradient boundary conditions. We approximate the spatially varying magnetic-field curvature by its field-line average, following an approach adopted in previous SOL turbulence studies~\cite{Killershanahan2020,Shanahan2020,Huslage2024,ShanahanFrontiers}. 
The high spatial resolution required to resolve the turbulent dynamics makes these simulations computationally demanding, with each case presented here requiring more than $500\,000$ CPU-hours.

In every case, a sheared radial electric field spontaneously develops near the abrupt connection-length transition (see Fig.~\ref{fig:Er_jump}a) whose magnitude scales inversely with the $L_{\parallel,f}$. We now show that this occurrence is robust, and can be described in terms of the asymptotic regions of uniform connection length.

Start by writing the plasma potential as $V_p=V_f+\Lambda(k_B/e)T_e$, where we separate the floating potential from the sheath potential drop~\cite[Sec.~2.6]{Stangeby}. Although the floating potential controls parallel currents, substantial current variations can generally be accommodated by relatively small changes in $V_f$~\citep[Ch.~17]{Stangeby}. The radial variation of $V_p$ in our asymptotic regions is therefore expected to be dominated by the electron-temperature gradient, as is commonly observed in the scrape-off layer~\citep[Ch.~17]{Stangeby}. Thus,

\begin{equation}
    E_x\approx -\Lambda\frac{k_B}{e}\frac{\mathrm{d}T_e}{\mathrm{d}x}, \label{eqn:Er_sheath}
\end{equation}

where the sheath factor $\Lambda$ is assumed approximately constant, neglecting its weak logarithmic dependence~\citep[Eq.~(5)]{tskhakaya2005boundary}. We further assume comparable radial scale lengths for the electron temperature ($T_e$), ion temperature ($T_i$), and density ($n$), such that $2L_p=L_{T_e}$, where $L_f=-(\mathrm{d}\ln f/\mathrm{d}x)^{-1}$. This is consistent with the simulations and expected unless the SOL is strongly conduction limited~\citep[Ch.~11]{Stangeby}.

To determine the equilibrium pressure gradient, we balance turbulent cross-field transport against parallel losses \cite{halpern2013theory,halpern2014theory,myra2015turbulent} in the continuity equation~\cite[Eq.~(1)]{Dudson2026},
\begin{align}
    \dfrac{\partial n}{\partial t} =     & -  \nabla \cdot \left[ n(\mathbf{b}v_{\parallel e}+\mathbf{v}_E+ \mathbf{v}_{de})\right], \label{eqn:continuity}
\end{align}
where explicit sources are neglected and $\mathbf{v}_E$, $\mathbf{v}_{de}$, and $v_{\parallel e}$ denote the $E\times B$ drift, diamagnetic drift, and parallel electron velocity, respectively. Decomposing the fields into equilibrium and smaller fluctuating components (denoted by tildes), and averaging over turbulent timescales and the $y$ and $z$ coordinates, gives

\begin{equation}
  \left\langle\nabla \cdot (nv_{\parallel e}\mathbf{b})\right\rangle = -\left\langle\nabla \cdot ( \tilde{n} \tilde{\mathbf{v}}_E)\right\rangle, \label{eqn:avrg_continuity}
\end{equation}
where the diamagnetic contribution vanishes in the electrostatic limit for a magnetic field uniform across the drift plane.

The parallel-loss term can be estimated from the sheath boundary condition. Using the marginal Bohm condition~\citep[Eq.~(39)]{riemann1991bohm}\citep[Ch.~2]{Stangeby} and assuming $V_f\ll T_e$, we obtain $\left\langle\nabla \cdot (nv_{\parallel e}\mathbf{b})\right\rangle\approx \bar{n}c_s/L_\parallel$, where $c_s$ is the ion sound speed. The perpendicular transport depends on the nonlinear correlation $\tilde n\widetilde{\mathbf{v}}_E$, requiring estimates of the fluctuation amplitudes $\tilde{n}$ and $\tilde{\phi}$, and dominant eddy scale, $k_x^{-1}$. Taking a fixed eddy aspect ratio $\mathcal{A}=k_z/k_x$, these can be obtained from the continuity and vorticity equations.

The vorticity equation~\citep[Eq.~(A7)]{Tocco2026} is given by,
\begin{align}
    \dfrac{\partial \Omega}{\partial t} =     & -  \nabla \cdot ( \Omega \mathbf{v}_E)  +  \nabla \cdot \mathbf{j}_\parallel  +  \nabla \cdot \mathbf{j}_\mathrm{d}, \label{eqn:vorticity}
\end{align}

where vorticity $\Omega = \nabla \cdot \left[(n_{i0} m_i/B^2) \nabla_\perp \left( \phi + p_i/q_i n_{i0} \right) \right]$ \citep[Eq.~(6)]{Dudson2026}, and $\mathbf{j}_\parallel$ and $\mathbf{j}_{\mathrm d}$ are the parallel and diamagnetic currents. Balancing linear growth against nonlinear advection, $\gamma\Omega\sim v_{Ex}\partial_x\Omega$, gives the saturated dominant outer scale amplitude~\cite{barnes2011critically,adkins2023scale,adkins2026asymptotic,ricci2006small,rogers2010low,Ricci2013},
\begin{equation}
     \frac{e\tilde{\phi}}{T}\sim\frac{1}{\mathcal{A}}\frac{1}{\rho^\star_s}\frac{\gamma L}{c_s}\frac{1}{(k_x L)^2}, \label{eqn:crit_balance}
\end{equation}
where $\rho_s^\star=\rho_s/L$ is the normalized sound gyroradius and $L$ some reference length. Similarly, the density continuity balance $\gamma \tilde{n}\sim v_{Ex}\partial_x\bar{n}$ yields
\begin{equation}
    \frac{\tilde{n}}{\bar{n}}\sim \frac{1}{k_x L_n}. \label{eqn:density_pert}
\end{equation}

For sufficiently long connection lengths, the dominant instability is the electrostatic interchange mode~\cite{rosenbluth1957stability,kesner2000interchange,Scott2005,Scott2008}. Consistently, the simulated fluctuations exhibit flute-like structures and an approximately $\pi/2$ phase shift between pressure and potential perturbations~\cite{Scott2005,Scott2008}. We therefore take
$\gamma\sim c_s/\sqrt{R_\mathrm{eff}L_p}$
~\citep[Eq.~(32)]{kesner2000interchange}\cite{ricci2006small},
where $R_\mathrm{eff}$ is the effective magnetic-field curvature radius.

Because the interchange instability is only weakly selective in perpendicular scale, the dominant eddy size must be determined by the nonlinear dynamics. 
We invoke critical balance~\cite{sridhar1994toward,nazarenko2011critical,barnes2011critically,nies2026saturation} to obtain the required eddy-size. Turbulent structures cannot remain coherent for longer than the characteristic parallel communication time to the sheath, limiting their perpendicular extent and amplitude. Similar arguments have also been used in the SOL literature \cite{Ricci2013,giacomin2020investigation,manz2025turbulence}. Balancing nonlinear advection with parallel transport in the vorticity equation, Eq.~(\ref{eqn:vorticity}), assuming $\tilde{\phi}$-dominated vorticity, and using Ohm's law $\tilde j_\parallel\sim k_\parallel\sigma_\parallel\tilde\phi$,
then gives 
\begin{equation}
    (k_xL)^2\sim \frac{(k_\parallel L)^2}{1+\mathcal{A}^2}\frac{1}{(\rho^\star_s)^2}\frac{c_s/L}{\gamma}\frac{\lambda_{e}}{L}\sqrt{\frac{m_i}{m_e}}, \label{eqn:mix_length}
\end{equation}
where $\lambda_e=v_{Te}\tau_e$ is the electron mean free path and
$\sigma_\parallel\sim ne^2\tau_e/m_e$ is the Spitzer conductivity~\citep[Eqs.~(2.7)--(2.8)]{Braginskii1965}, with electron collision time $\tau_e$~\citep[Eq.~(2.5e)]{Braginskii1965}.

Combining Eqs.~(\ref{eqn:crit_balance}), (\ref{eqn:density_pert}), and (\ref{eqn:mix_length}) with the averaged continuity balance, Eq.~(\ref{eqn:avrg_continuity}), yields $L_p\sim\alpha_pL_\parallel$  where 

\begin{equation}
    \alpha_p\sim(1+\mathcal{A}^2)^{1/3}{(\rho_s^{\star})}^{2/3}\left(\frac{L^2}{R_\mathrm{eff}\lambda_{e}}\right)^{1/3}\left(\frac{m_e}{m_i}\right)^{1/6}. \label{eqn:Lp_scale}
\end{equation}

\begin{figure}
    \centering
    \includegraphics[width=0.9\linewidth]{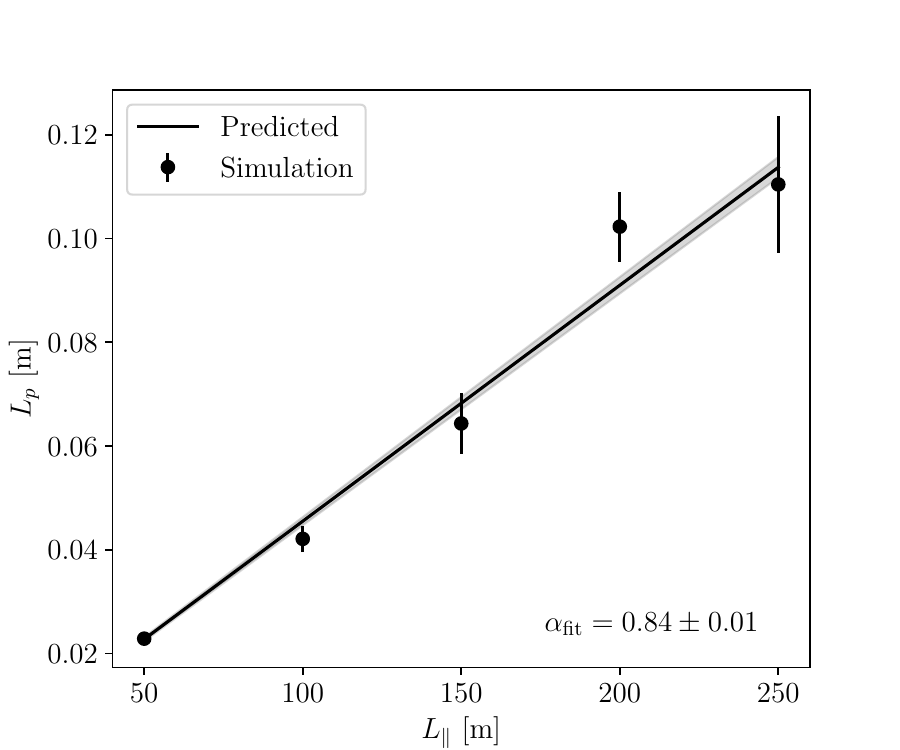}
    \caption{\textbf{Predicted and simulated $L_p$ for different connection lengths $L_\parallel$.} Comparison of the pressure gradient in simulations (scatter) with the prediction in Eq.~(\ref{eqn:Lp_scale}). The theoretical prediction uses characteristic simulation parameters as gathered in Table~\ref{tab:scaling_parameters}, as well as a fitting factor $\alpha_\mathrm{fit}=0.81\pm0.01$. Error bars indicate the variation of $L_p$ over the final time-averaging window in the simulations.}
    \label{fig:Lp}
\end{figure}

\begin{figure*}
    \centering
    \includegraphics[width=0.9\linewidth]{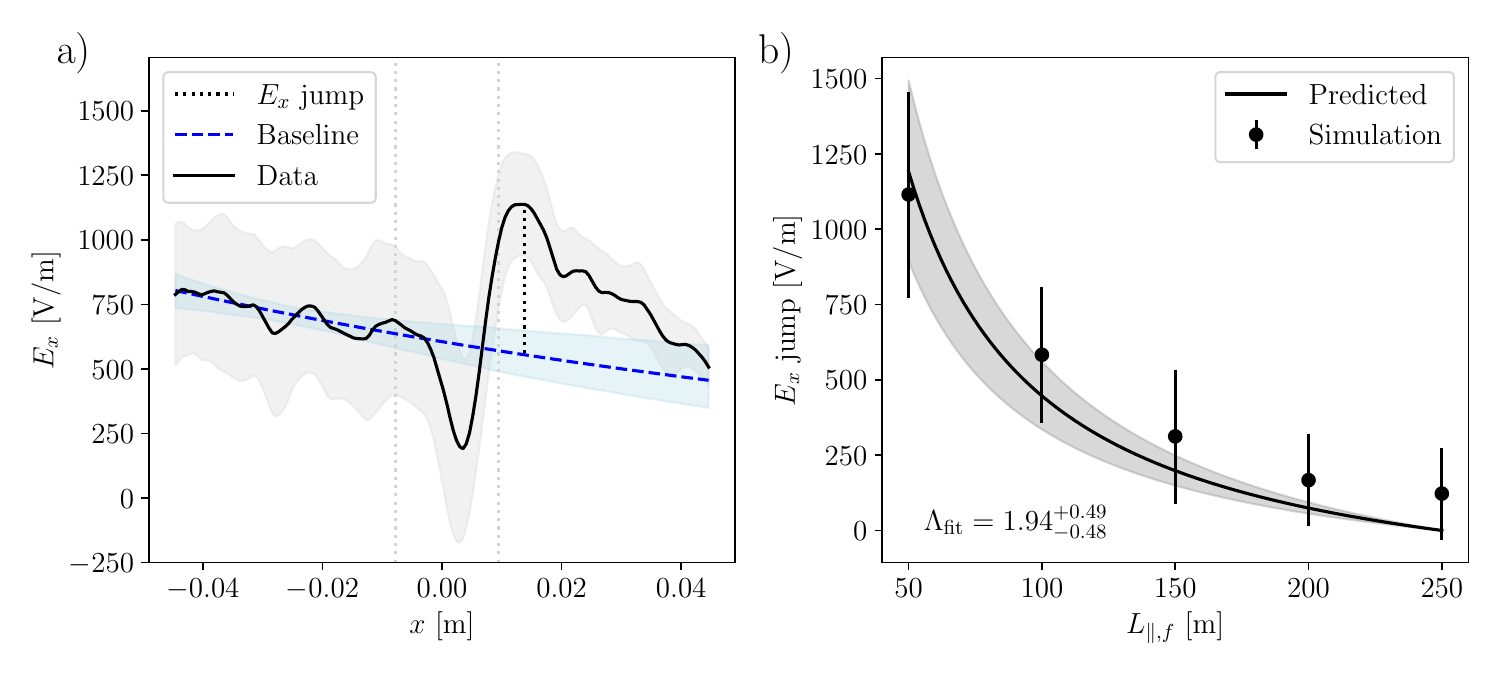}
    \caption{\textbf{Jump in the radial electric field.} (a) Example of the simulated $E_x$ jump for $L_{\parallel,f}=100$ m. (b) Comparison of the measured jump (scatter) with the theoretical scaling (black line) as a function of the final connection length $L_{\parallel,f}$. The measured jump is the maximum electric field relative to a baseline obtained by fitting an exponential to $E_x$ to the left of the transition. Uncertainties arise from temporal variability (gray shading in a) and the baseline fit (blue shading). The prediction uses a single fitting parameter, $\Lambda_\mathrm{fit}=1.94\pm0.49$, comparable to the sheath factor $\Lambda\approx2.3$ for ions with an MHD-like adiabatic index of $5/3$~\citep[Eq.~(5)]{tskhakaya2005boundary}.}
    \label{fig:Er_jump}
\end{figure*}

The resulting linear dependence $L_p\propto L_\parallel$ captures the principal effect of connection length on cross-field transport: shorter connection lengths limit turbulent development before parallel losses to the sheath, reducing fluctuation amplitudes and scales, and requiring steeper equilibrium gradients. Figure~\ref{fig:Lp} compares the prediction with simulations spanning different $L_\parallel$. Using characteristic simulation parameters (Table~\ref{tab:scaling_parameters}), a single order-unity fitting factor, $\alpha_\mathrm{fit}\approx0.81$, reproduces the observed $L_p$. It must be noted that shorter connection lengths and steeper gradients  favor drift-wave instabilities~\cite{Scott2005,Ricci2011,Hasegawa1983, mosetto2013turbulent}, which are neglected here and may modify the quantitative scaling, although not the underlying mechanism.

Together with Eq.~(\ref{eqn:Er_sheath}), the dependence $L_p\propto L_\parallel$ implies different asymptotic radial electric fields on either side of the connection-length transition. Connecting these asymptotic regimes necessarily produces a finite electric-field jump, independently of the detailed transition-region dynamics. A sheared electric field is therefore a robust consequence of an abrupt connection-length change. This sheared electric field arises even where the temperature is nearly uniform across the transition. This shows that a self-consistent determination of the shear around the transition region would require including the current balance governing the floating potential~\cite{Loizu2017}, and also the parallel dynamics~\cite[Part~II]{Stangeby}. A similar scaling approach is possible in that case, but is left to a future publication.

Figure~\ref{fig:Er_jump} compares the predicted asymptotic difference
$\Delta E_{x,\mathrm{asym}}
=E_{x,\mathrm{asym}}(L_{\parallel,f})
-E_{x,\mathrm{asym}}(L_{\parallel,i})$
with the simulated electric-field variation. Exact agreement is not expected, since the asymptotic prediction neither resolves the transition structure nor uniquely determines the local jump measure. Nevertheless, the agreement in magnitude and connection-length dependence, with a fitted $\Lambda$ consistent with the expected sheath factor~\citep[Eq.~(5)]{tskhakaya2005boundary}, supports the proposed mechanism.

The dynamical significance of this electric-field variation can be quantified by the resulting $E\times B$ shear~\cite{Biglari1990,Terry2000}, which we characterize by $\Theta=\gamma_E/\gamma$, where
$\gamma_E=B^{-1}\mathrm{d}E_x/\mathrm{d}x$
is the shearing rate and $\gamma$ the linear instability growth rate. Values $\Theta\sim O(1)$ indicate that the shear can disrupt the underlying turbulence. Leaving the fully self-consistent treatment aside, we assume that the electric field changes between its asymptotic values over the connection-length transition width $\Delta x$, and taking a reference $\gamma$ to the left of the transition, we estimate

\begin{equation}
    \Theta\sim\Lambda (\rho_s^\star)^{2/3}\left(\frac{R_\mathrm{eff}}{L}\right)^{2/3}\left(\frac{L}{L_{\parallel,i}}\right)^{1/2}\frac{1}{\Delta x/L}\frac{1-f_\parallel}{f_\parallel}, \label{eqn:shearing_rate}
\end{equation} 

where $f_\parallel=L_{\parallel,f}/L_{\parallel,i}$ and weak dependencies on collisionality and mass ratio (powers of $1/6$ or $1/12$) have been omitted.

For the simulations considered here, this gives
$\Theta\sim1.5~(1/f_\parallel-1)$.
Thus, sufficiently strong connection-length variations naturally generate shear comparable to the instability growth rate, even though shear was not included in the turbulence scaling used to predict the equilibrium gradients. The effect is expected and predicted to weaken at higher magnetic fields, owing to the reduced $E\times B$, and under stronger curvature-driven transport, which relaxes equilibrium gradients. The prediction cannot, however, be extrapolated to arbitrarily narrow transitions. 
The scale separation
\begin{equation}
    k_x\rho_s\gg (\rho_s/L_\parallel)\mathrm{d}L_\parallel/\mathrm{d}R\gg \rho_s/L_p,\label{eqn:length_scale_ordering}
\end{equation}
reflected in simulations with $k\rho_s\sim10^{-2}$ and $kL_p\sim10^1$, eventually fails, requiring additional physics to determine the minimum transition width. If the induced radial electric field shear becomes sufficiently large, the shearing estimate will also fail, as the Kelvin-Helmholtz (KH) instability kicks in~\cite{Myra2016}. This could produce strong filamentary (blob) transport~\cite{Walkden2019} which is not explicitly treated in this analysis.

\begin{table}
    \centering
    \caption{Reference parameters used for the scaling estimates.}
    \label{tab:scaling_parameters}
    \begin{tabular}{lcc}
        \hline
        Parameter & Symbol & Value \\
        \hline
        Effective major radius & $R_{\mathrm{eff}}$ & $5.5~\mathrm{m}$ \\
        Transition width & $\Delta x$ & $3.75\times10^{-3}~\mathrm{m}$ \\
        Initial connection length & $L_{\parallel,i}$ & $250~\mathrm{m}$ \\
        Domain radius & $R_{\mathrm{dom}}$ & $0.15~\mathrm{m}$ \\
        Temperature & $T_e,~T_i$ & $35~\mathrm{eV}$ \\
        Density & $n$ & $4\times10^{18}~\mathrm{m}^{-3}$ \\
        Magnetic field & $B$ & $2.5~\mathrm{T}$ \\
        \hline
    \end{tabular}
\end{table}
The simulations are limited by a simplified treatment of the shallow angle sheath boundary condition~\cite{Geraldini2024, Loizu2012} and simplified geometry of the W7-X SOL. In doing so, we ignored the 3D variation of equilibrium field quantities (such as the field curvature or $B$) and the more complex arrangement of connection length transitions (including $z$ dependence)~\cite[Fig.~1]{Kriete2023}. The presence of a radial electric field shear induced by a sharp connection length transition has been shown to be robust, however, and is therefore expected to persist when including these effects. This must therefore be taken into consideration as an additional mechanism in any electric field analysis of the SOL. 

\input{sections/acknowledgements}

\bibliography{master}


\end{document}

%% file: preamble.tex
\usepackage{amsthm}
\usepackage{mathtools}
\usepackage{physics}
\usepackage{xcolor}
\usepackage{graphicx}
\usepackage[left=23mm,right=13mm,top=35mm,columnsep=15pt]{geometry} 
\usepackage{adjustbox}
\usepackage{placeins}
\usepackage[T1]{fontenc}
\usepackage{lipsum}
\usepackage{url}
\usepackage{csquotes}
\usepackage{subfig}
\usepackage{url}

%% file: sections/acknowledgements.tex
\section*{Acknowledgments} \label{sec:acknowledgments}
The first author (BS) would like to acknowledge the work of the BOUT++ \& Hermes-3 developer community. BOUT++ is an open-source framework available at \href{https://boutproject.github.io}{boutproject.github.io}. Hermes-3 is open source, and found at \href{https://github.com/boutproject/hermes-3}{github.com/boutproject/hermes-3}. The second author (ER) would like to thank fruitful discussion with J. Loizu and R. Nies.
The input files for the simulations shown herein are available at \href{https://github.com/bshanahan/papers/PRL2026}{github.com/bshanahan/papers/ShanahanRodriguez2026}. 